\documentclass[twocolumn,superscriptaddress]{revtex4-1}

\usepackage{graphicx}
\usepackage{epstopdf}
\usepackage{color}
\usepackage{siunitx}

\DeclareSIUnit{\gauss}{G}

\begin{document}


\title{Systematic optimization of laser cooling of dysprosium}


\author{Florian M\"uhlbauer}
\email[]{muehlbau@uni-mainz.de}
\affiliation{Institut f\"ur Physik, AG QUANTUM, Johannes Gutenberg-Universit\"at Mainz, 55122 Mainz, Germany}

\author{Niels Petersen}
\affiliation{Institut f\"ur Physik, AG QUANTUM, Johannes Gutenberg-Universit\"at Mainz, 55122 Mainz, Germany}
\affiliation{Graduate School Materials Science in Mainz, Staudingerweg 9, 55128 Mainz, Germany}

\author{Carina Baumg\"artner}
\affiliation{Institut f\"ur Physik, AG QUANTUM, Johannes Gutenberg-Universit\"at Mainz, 55122 Mainz, Germany}

\author{Lena Maske}
\affiliation{Institut f\"ur Physik, AG QUANTUM, Johannes Gutenberg-Universit\"at Mainz, 55122 Mainz, Germany}

\author{Patrick Windpassinger}
\affiliation{Institut f\"ur Physik, AG QUANTUM, Johannes Gutenberg-Universit\"at Mainz, 55122 Mainz, Germany}
\affiliation{Graduate School Materials Science in Mainz, Staudingerweg 9, 55128 Mainz, Germany}


\date{\today}

\begin{abstract}
We report on an apparatus for cooling and trapping of neutral dysprosium. We characterize and optimize the performance of our Zeeman slower and 2D molasses cooling of the atomic beam by means of Doppler spectroscopy on a \SI{136}{\kilo\hertz} broad transition at \SI{626}{\nano\meter}. Furthermore, we demonstrate the characterization and optimization procedure for the loading phase of a magneto-optical trap (MOT) by increasing the effective laser linewidth by sideband modulation. After optimization of the MOT compression phase, we cool and trap up to $10^9$ atoms within 3 seconds in the MOT at temperatures of \SI{9}{\micro\kelvin} and phase space densities of $1.7 \cdot 10^{-5}$, which constitutes an ideal starting point for loading the atoms into an optical dipole trap and for subsequent forced evaporative cooling. 

\end{abstract}


\maketitle

\section{Introduction \label{sec:intro}}

\begin{figure*}
  \includegraphics[width=\textwidth]{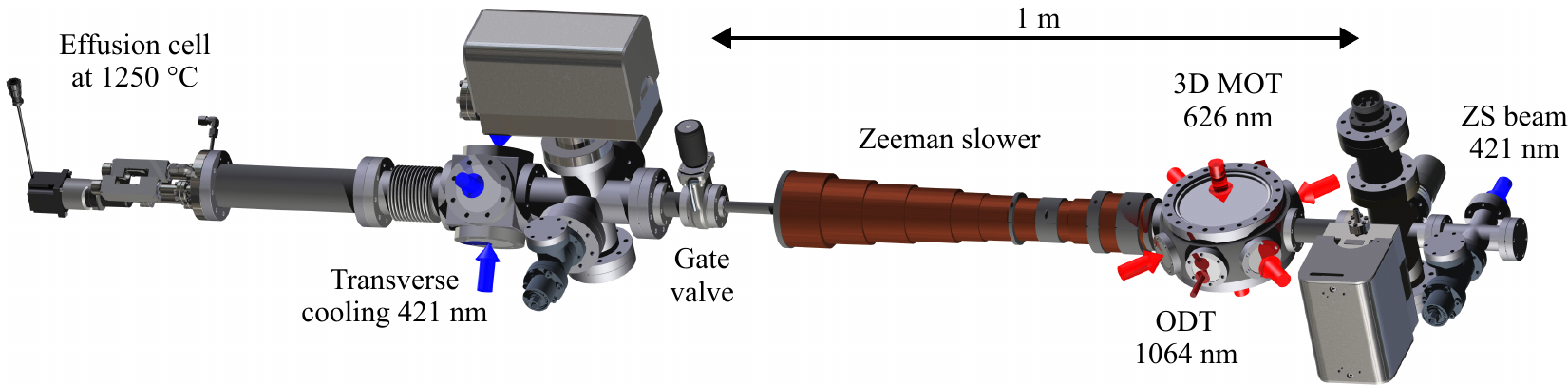}
\caption{CAD drawing of the vacuum system. In the oven chamber on the left hand side, Dy atoms are evaporated in an effusion cell and transversally cooled in a 2D molasses setup. After longitudinal deceleration inside the ZS the atoms are captured by the 3D MOT in the main chamber.}
\label{fig1:vacuum} 
\end{figure*}

Quantum gases composed of neutral atoms possessing large permanent magnetic ground-state dipole moments have proven to be a very successful platform for exploring dipolar physics in the quantum regime. These systems have led to the discovery of new quantum phases and quantum effects, for example the realization of the extended Bose-Hubbard model \cite{Baier2016}, the observation of the roton mode \cite{Chomaz2017} and droplet formation \cite{Ferrier2016,Ferrier2016liquid,Kadau2016,Schmitt2016,Chomaz2016}. Similar to these effects, which arise when the long-range, anisotropic magnetic dipole-dipole interaction (DDI) \linebreak dominates the system, other exotic states of quantum matter are predicted for dipolar bosonic quantum gases in optical lattices, e.g. checkerboard or supersolid phases \cite{Lahaye2009}. \\
Due to these fascinating prospects, the interest in laser cooling of lanthanide atoms with large permanent magnetic dipole moments has grown rapidly within the last few years \cite{Mcclelland2006,Berglund2007,Berglund2008,Leefer2010,Youn2010,Youn2010mot,Sukachev2010,Frisch2012,Maier2014,Miao2014,Dreon2017,Lucioni2017,Ilzhofer2017}, as laser cooled atomic samples constitute the prerequisite for producing strongly magnetic degenerate quantum gases \cite{Lu2011bec,Lu2012dfg,Aikawa2012bec,Aikawa2014dfg,tang2015}. The generation of cold ensembles of lanthanide atoms in a magneto-optical trap (MOT) at temperatures of a few micro Kelvin has proven to be challenging \cite{Berglund2008,Lu2011bec}. However, a very successful and convenient approach was demonstrated \cite{Frisch2012,Maier2014}, by making use of strong optical transitions in the blue spectral range for precooling of the atoms inside a Zeeman slower (ZS) before capturing them in a narrow-line MOT. Here, a slow atom beam is produced by using cooling transitions with a linewidth of several tens of Megahertz, whereas the MOT is operated on transitions with natural linewidths on the order of \SI{100}{\kilo\hertz}, resulting in Doppler temperatures below \SI{5}{\micro\kelvin}, which allows to directly transfer the atoms from the MOT into an optical dipole trap (ODT). Moreover, the interplay of radiation pressure cooling on narrow line transitions and gravitational forces lead to spin polarization of the atomic sample in the MOT \cite{Berglund2008,Frisch2012,Maier2014,Dreon2017,Lu2011bec}, which allows for describing the system with a two-level model. \\
Dysprosium possesses the highest permanent magnetic ground-state dipole moment in the periodic table alongside terbium, which makes it an ideal choice for performing quantum gas experiments with dipolar atoms. It belongs to the group of lanthanide elements whose characteristic open f-shell electron configuration \linebreak $[$Xe$]$4f$^{10}$6s$^2$ with spin S = 2, orbital angular momentum L = 6 and total angular momentum J = 8 gives rise to its high magnetic moment of 10 Bohr magnetons ($\si{\micro} \sim 10\, \si{\micro} _{\mathrm B}$). Compared to alkali atoms, whose magnetic moments are on the order of only $\si{\micro} \sim 1\, \si{\micro} _{\mathrm B}$, the DDI in ultra-cold gases of dysprosium is about 100 times stronger. However, due to its low vapor pressure, temperatures exceeding \SI{1000}{\celsius} are necessary to evaporate a sufficient amount of atoms from the dysprosium source material. This leads to high initial atomic velocities of several hundred meters per second, which have to be reduced considerably before trapping of the atoms can be achieved. For precooling and Zeeman slowing we employ a strong J = 8 $\rightarrow$ J$^{\prime}$ = 9 transition at \SI{421}{\nano\meter} \cite{Leefer2009,Lu2011,Martin1978} with a natural linewidth of $\mathrm\Gamma_{421}=2\pi \cdot \SI{32}{\mega\hertz}$ and a saturation intensity of $\mathrm I_{421}^{\mathrm{sat}}=\SI[per-mode=symbol]{56}{\milli\watt\per\square\centi\meter}$. This transition is also used for absorption imaging of the atomic cloud. Even though this is not a closed cycling transition, it is well suited due to its adequate branching ratio to the ground state \cite{Leefer2010,Youn2010}. For the MOT, we use a J~= 8 $\rightarrow$ J$^{\prime}$ = 9 transition at \SI{626}{\nano\meter} \cite{Martin1978,hogervorst1978,Gustavsson1979} with a natural linewidth of $\mathrm\Gamma_{626}=2\pi \cdot \SI{136}{\kilo\hertz}$, corresponding to a Doppler temperature of \SI{3.2}{\micro\kelvin}, and a saturation intensity of $\mathrm I_{626}^{\mathrm{sat}}=\SI[per-mode=symbol]{72}{\micro\watt\per\square\centi\meter}$. It is a closed transition as decay into lower energy levels other than the ground state are forbidden by parity and angular momentum selection rules, thus making a repump laser for the MOT unnecessary. \\
This paper concentrates on the systematic optimization of different laser cooling parameters for dysprosium. The main part of the manuscript is organized in the following way: First we present our experimental setup (Sec. \ref{sec:exp}), designed for cooling and trapping of dysprosium atoms and give detailed descriptions of the vacuum system (Sec. \ref{sec:vacuum}) and the laser system (Sec. \ref{sec:laser}). In sections \ref{sec:expro} and \ref{sec:method}, we briefly summarize the experimental procedure and the methods employed for characterization and parameter optimization measurements, respectively. In section \ref{sec:result} we show the results obtained for the transverse 2D molasses cooling (Sec. \ref{sec:tc}), the Zeeman slower (Sec. \ref{sec:zs}) and for the MOT loading and compression phase (Sec. \ref{sec:mot}). 

\section{Experimental setup \label{sec:exp}}
Our experimental setup consists of an ultra-high vacuum chamber and a dedicated laser system to generate tunable, frequency stabilized laser radiation necessary for laser cooling of dysprosium. Experimental sequences are generated by a LabVIEW based computer program which controls a real-time sequence processor. In this section, we will present the vacuum system and the laser system in detail.

\subsection{Vacuum system \label{sec:vacuum}}
To reduce the influence of undesired magnetic stray fields on the Dy atoms, the vacuum system is assembled exclusively from anti-ferromagnetic stainless steel components. Figure \ref{fig1:vacuum} shows a CAD drawing of the vacuum system. It consists of two parts: the oven chamber and the main chamber. Both parts are separated by a gate valve allowing us to vent them individually, e.g. for refilling of the oven (SVT Associates, Inc., SVTA-DF-10-450, 10cc), which is used for evaporating the dysprosium source material. It is attached to the main body of the oven chamber by means of a flexible bellow, thus allowing for adjustment of the atomic beam. A UHV 6-way cube equipped with four DN 63 viewports, provides excellent optical access for transverse cooling (TC) of the atomic beam. We have installed an aperture of \SI{4}{\milli\meter} diameter in front of the TC cube to collimate the atomic beam and to protect the vacuum viewports from being coated with dysprosium. The oven chamber is pumped by a \SI[per-mode=symbol]{75}{\liter\per\second} ion pump, which maintains pressures below $8\cdot 10^{-10}\, \si{\milli\bar}$ during full oven operation. \\
Downstream of the gate valve, the ZS connects the oven chamber and the main chamber. The ZS coil assembly encloses a \SI{750}{\milli\meter} long DN 16 vacuum pipe, which, due to its low conductance, acts as a differential pumping stage and is able to sustain pressure differences of at least two orders of magnitude. The main chamber consists of a spherical octagon (Kimball Physics, MCF800M-SphOct-G2C8-A), which provides all necessary optical access for the 3D MOT, absorption imaging and the optical dipole trap. It is attached to a supporting 6-way-cross, which incorporates a \SI[per-mode=symbol]{45}{\liter\per\second} ion pump and a titanium sublimation pump, achieving vacuum pressures as low as $4\cdot 10^{-11}\, \si{\milli\bar}$. We have placed a right-angle prism mirror with aluminum coating inside the vacuum chamber to deflect the \SI{421}{\nano\meter} ZS laser beam towards the oven chamber, counter-propagating to the thermal atomic beam.

\begin{figure}
  \includegraphics{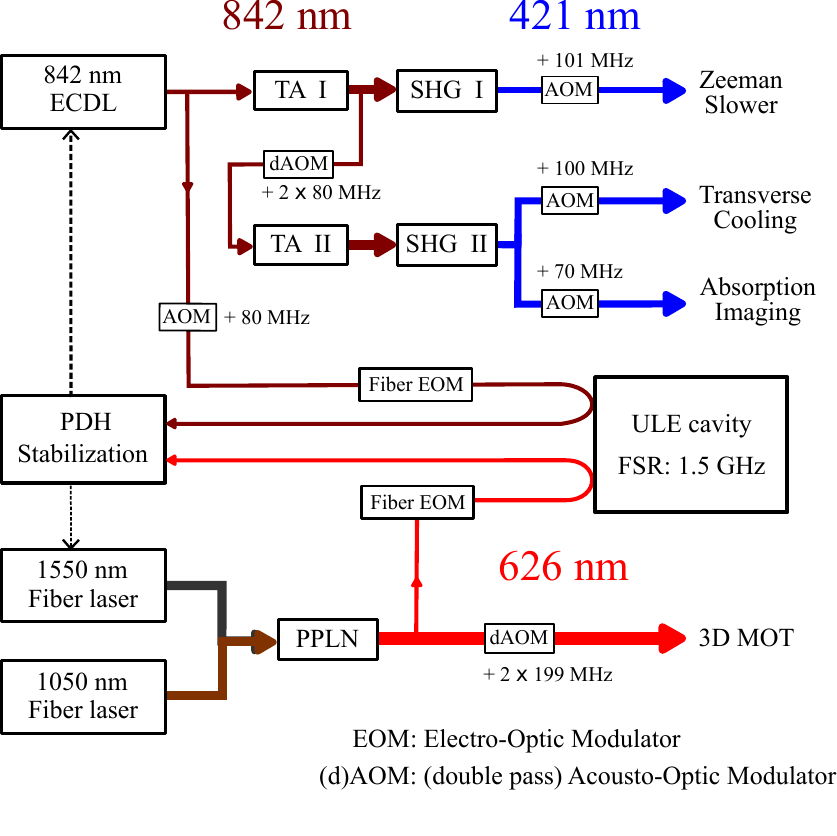}
\caption{Overview of the laser system. Starting with one \SI{842}{\nano\meter} master ECDL and subsequent power amplification (TA) the \SI{421}{\nano\meter} light is generated in two frequency doubling cavities (SHG), while the \SI{626}{\nano\meter} light is generated by sum frequency mixing of the output light of two fiber lasers in a PPLN crystal. Frequencies are shifted via AOMs and EOMs with the fundamental lasers being stabilized via PDH offset-sideband locking to a ULE reference cavity.}
\label{fig2:laser} 
\end{figure}

\subsection{Laser system \label{sec:laser}}
The laser system is depicted schematically in Fig. \ref{fig2:laser}. We produce the \SI{421}{\nano\meter} laser radiation for TC, ZS and absorption imaging by second harmonic generation (SHG) in two homebuilt frequency doubling units, each one routinely delivering up to \SI{220}{\milli\watt} of \SI{421}{\nano\meter} light. These units consist of a temperature stabilized beta-barium borate (BBO) crystal placed inside a power enhancing bow-tie resonator, which is length stabilized onto the fundamental laser wavelength of \SI{842}{\nano\meter}. We use a commercial \SI{842}{\nano\meter} external cavity diode laser (ECDL), which is seeding two homebuilt tapered amplifiers (TA), providing up to \SI{2}{\watt} of output power each. After the first TA, a small fraction of the \SI{842}{\nano\meter} light is separated and frequency shifted by an acousto-optical modulator (AOM), set up in double pass configuration, before it is used to seed the second TA. In this way, we are able to generate red-detuned \SI{421}{\nano\meter} light for the ZS and near-resonant \SI{421}{\nano\meter} light for TC and absorption imaging by using only one fundamental laser. The fundamental \SI{842}{\nano\meter} diode laser is frequency stabilized to an ultra-stable, ultra-low expansion (ULE) reference cavity with a free spectral range of \SI{1.5}{\giga\hertz} by offset-sideband Pound-Drever-Hall (PDH) locking \cite{Thorpe2008}. To this end, we modulate the diode laser current with a frequency of \SI{20}{\mega\hertz} and use a fiber coupled electro-optical phase-modulator (EOM) to produce additional sidebands in the frequency range of \SI{750}{\mega\hertz} to \SI{1.5}{\giga\hertz}. In this way, we can produce PDH error signals for laser stabilization at any frequency between two adjacent ULE cavity resonances, thus allowing us to span the frequency difference between our ULE cavity and the resonance frequencies of all relevant dysprosium isotopes. \\
The orange \SI{626}{\nano\meter} laser light for the 3D MOT is produced in a sum frequency mixing setup following the approach demonstrated by Wilson et al. \cite{Wilson2011}. The output light of two fiber-laser-amplifier systems (NKT Photonics, Koheras BoostiK E15 and Koheras BoostiK Y10 PM) with wavelengths of \SI{1550}{\nano\meter} and \SI{1050}{\nano\meter} is spatially overlapped and focused through a periodically poled lithium niobate (PPLN) crystal, where \SI{626}{\nano\meter} laser radiation is produced by nonlinear frequency conversion. The crystal is placed inside a temperature stabilized oven to achieve quasi-phase matching and ensure optimal frequency conversion. For a maximum power of \SI{5}{\watt} from both fiber lasers, this system generates up to \SI{2.2}{\watt} of \SI{626}{\nano\meter} output power. We also use offset-sideband locking to enable easy switching between the different dysprosium isotopes. The local oscillator at \SI{11}{\mega\hertz} for the PDH and the variable drive signal \linebreak (\SI{750}{\mega\hertz} to \SI{1.5}{\giga\hertz}) for the frequency offset are combined on a power splitter. The PDH error signals from the \SI{626}{\nano\meter} light, which we couple to the ULE cavity, is fed back to the \SI{1550}{\nano\meter} fiber laser, while the \SI{1050}{\nano\meter} fiber laser remains free-running.

\begin{figure*}
  \includegraphics[width=\textwidth]{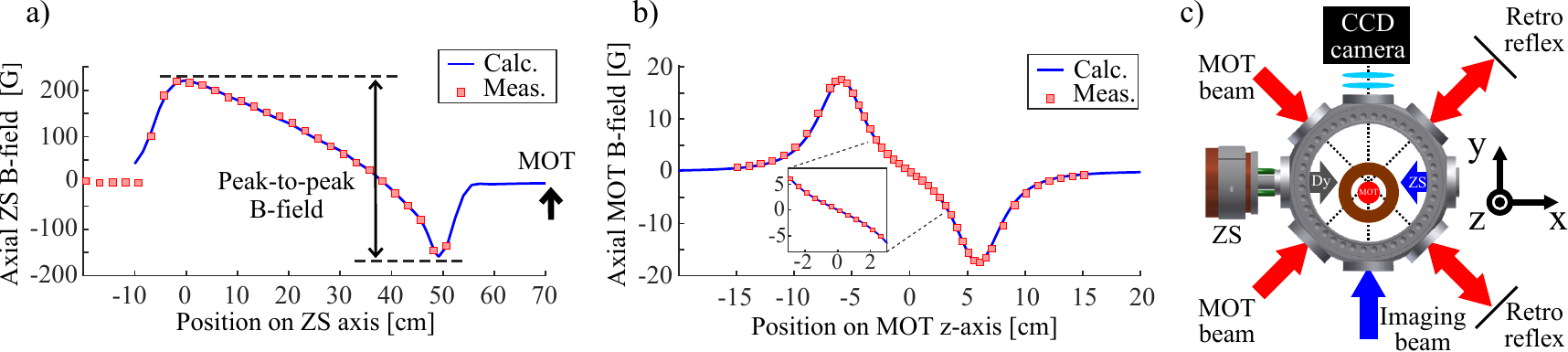}
\caption{(a) Axial magnetic field profile of the Zeeman slower (along the x-axis), operated at \SI{9.5}{\ampere} (simulation data from the manufacturer). The MOT position is marked by the arrow at \SI{71}{\centi\meter}. (b) Axial magnetic field profile generated by the anti-Helmholtz coils (along the z-axis). The inset shows the magnetic field in the MOT region between $\pm$\SI{2}{\centi\meter}. (c) Schematic view of the MOT setup and the laser beam geometry for the ZS and the imaging system in the main chamber. The MOT coils (depicted as a copper ring) are displaced from the atomic beam axis by \SI{5}{\milli\meter} to reduce the influence of the ZS beam on the atoms trapped in the MOT. Gravity is pointing along the negative z-axis.}
\label{fig3:mag_fields}
\end{figure*}

\section{Experimental procedure \label{sec:expro}}
We operate the dual-filament high-temperature effusion cell at \SI{1230}{\celsius} (\SI{1250}{\celsius}) in the main (tip) filament to produce a thermal atomic beam of dysprosium with a peak velocity of about \SI[per-mode=symbol]{480}{\meter\per\second}. First, the atoms are laser cooled in transverse direction in order to reduce the beam divergence and hence increase the flux of atoms through the narrow ZS pipe into the main chamber. This is achieved in a 2D molasses cooling setup, where we employ the strong transition at \SI{421}{\nano\meter}. We use two elliptical laser beams, set up in retro-reflex configuration, with $1/\mathrm e^2$ diameters of \SI{19}{\milli\meter} $\times$ \SI{5}{\milli\meter} to obtain a large overlap volume with the atomic beam. Laser cooling is typically achieved at a detuning of $\mbox{-1}$~$\mathrm\Gamma_{421}$ and a laser power of \SI{70}{\milli\watt} per transverse cooling beam, which corresponds to a mean intensity of 1.7~$\mathrm I_{421}^{\mathrm{sat}}$. \\
Next the atoms are longitudinally decelerated in the ZS to peak velocities of about \SI[per-mode=symbol]{24}{\meter\per\second} before they enter the main chamber. We use a ZS in spin-flip configuration to limit the maximum absolute magnetic field strength to less than \SI{250}{\gauss}, which reduces the power consumption of the coil assembly to about \SI{35}{\watt}. It consists of 11 individual coils, that are contacted in series to be supplied from the same current source with typically \SI{9.5}{\ampere}. The last coil at the end of the ZS is contacted opposite to the one before to reduce the residual magnetic field leaking out of the ZS into the proximity of the MOT. The theoretically calculated and measured magnetic field profiles of the commercially produced ZS (OSWALD Elektromotoren GmbH) are shown in Fig. \ref{fig3:mag_fields}a. To dissipate the heat produced from the coil assembly, we have implemented water-cooling between the vacuum chamber and the  ZS pipe (\SI{40}{\milli\meter} inner diameter). With a water flow of about \SI[per-mode=symbol]{10}{\liter\per\minute} and a water temperature of \SI{18}{\celsius}, we measure maximum temperatures of less than \SI{25}{\celsius} during ZS operation on the surface of the coils. The laser cooling is achieved with typically \SI{110}{\milli\watt} of circularly polarized \SI{421}{\nano\meter} light. The cooling beam is focused towards the oven aperture and has a beam diameter of \SI{14}{\milli\meter} (1/e$^2$) at the MOT position. \\
In the main chamber, slow atoms are captured by a six-beam 3D MOT, set up in retro-reflex configuration, which is operated on the closed, narrow line \SI{626}{\nano\meter} transition. In order to reduce the influence of the \SI{421}{\nano\meter} ZS beam passing through the MOT region, we have mounted the anti-Helmholtz coils \SI{5}{\milli\meter} off-center with respect to the atomic beam axis (see Fig. \ref{fig3:mag_fields}c). Both coils have a mean diameter of \SI{68}{\milli\meter} and are placed in a distance of \SI{118}{\milli\meter} to each other. Figure \ref{fig3:mag_fields}b shows the measured and calculated axial magnetic field profile produced by the MOT coils. In the center, this setup provides a linear magnetic field gradient of about \SI{40}{\milli\meter} spatial extent. The \SI{626}{\nano\meter} laser beams are expanded to large diameters of \SI{17}{\milli\meter} (1/e$^2$) to increase the MOT trapping volume. We increase the capture rate and the loading efficiency by artificial broadening of the \SI{626}{\nano\meter} laser linewidth, which will be discussed in Sec. \ref{sec:mot}.

\section{Methods \label{sec:method}} 
To characterize the velocity distribution of the atomic beam, we perform velocity sensitive fluorescence spectroscopy \cite{Prodan1982} on the \SI{626}{\nano\meter} transition. Due to its narrow linewidth of \SI{136}{\kilo\hertz}, we can achieve higher resolution by utilizing this transition compared to the broader \SI{421}{\nano\meter} transition. To this end, a collimated, circularly polarized \SI{626}{\nano\meter} laser beam is sent through the main chamber to intersect with the atomic beam at the MOT position. While scanning the laser frequency across the atomic resonance, we collect the fluorescence photons with a photomultiplier mounted perpendicular to both the spectroscopy beam and the atomic beam. The relative angle between the probe beam and the atomic beam is henceforth denoted as $\mathrm\alpha$. Transverse velocity distributions can be measured if the probe beam is aligned to $\mathrm\alpha = \SI{90}{\degree}$, whereas longitudinal velocity distributions can be accessed for $\mathrm\alpha \neq \SI{90}{\degree}$. In the latter case, the Doppler shift of the atomic resonance frequency is given by $\omega_D=2\pi f_D=kv\cos(\mathrm\alpha)$. Here, $k$ and $v$ denote the modulus of the spectroscopy beam wavevector and the longitudinal velocity component of the atoms, respectively. For frequency to velocity conversion, we determine the zero-velocity reference frequency from a spectrum recorded at $\mathrm\alpha = \SI{90}{\degree}$. Due to the nonlinear conversion factor, velocities can be determined more precisely for smaller values of $\mathrm\alpha$, since uncertainties in its determination become less important for small angles. We use this technique to benchmark the performance of our ZS, as will be described in Sec. \ref{sec:zs}. \\
For characterization of the MOT, we use time-of-flight absorption imaging. Therefore, a large, collimated \linebreak \SI{421}{\nano\meter} laser beam with an $1/\mathrm e^2$ diameter of \SI{30}{\milli\meter} is aligned along the y-axis (perpendicular to the direction of gravity, see Fig. \ref{fig3:mag_fields}c) and covers enough space to image the atomic cloud for time-of-flights up to \SI{30}{\milli\second}. While imaging, we apply a weak magnetic field along the imaging axis to define a quantization and thus enhance the light-atom interaction cross-section due to favorable Clebsch-Gordan coefficients in the stretched Zeeman states. Typically, we use a laser power of \SI{0.4}{\milli\watt} for imaging, corresponding to an intensity of 0.001~$\mathrm I_{421}^{\mathrm{sat}}$ and detune the frequency by 1~$\mathrm\Gamma_{421}$ to avoid saturation effects during imaging.

\section{Results \label{sec:result}}
In this section, we first summarize the characterization measurements of the TC, while the main focus will be the results of our optimization procedure for the ZS and the MOT. All measurements have been performed with the bosonic isotope $^{162}$Dy, and have, for the ZS spectroscopy measurements, also been confirmed for the fermionic isotope $^{163}$Dy and the most abundant bosonic isotope $^{164}$Dy.

\subsection{Transverse cooling \label{sec:tc}}

\begin{figure}
  \includegraphics{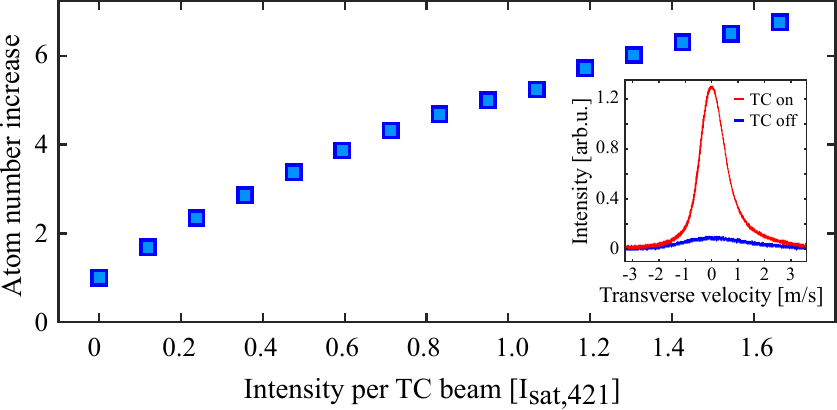}
\caption{Characterization of the TC setup. Atom number increase in the main chamber as a function of the TC beam intensity. Inset: Comparison of atom spectra for TC beams blocked (blue) and a TC beam intensity of 1.7~$\mathrm I_{421}^{\mathrm{sat}}$ (red)}
\label{fig4:tc_opt} 
\end{figure}

To characterize the transverse 2D molasses cooling in our setup, we perform fluorescence spectroscopy of the atomic beam, as described in Sec. \ref{sec:method}. To this end, we align the probe beam to $\mathrm\alpha = \SI{90}{\degree}$ and record atom spectra for different TC beam intensities (example spectra see inset of Fig. \ref{fig4:tc_opt}). For each spectrum we determine the area below the peak, which we take as a measure for the atom number, and compare it to the area when the TC beams are blocked. Figure \ref{fig4:tc_opt} shows a plot of relative atom numbers as a function of the TC beam intensity. We measure a continuous increase of atoms arriving in the main chamber with increasing TC beam intensity up to a mean intensity of 1.7~$\mathrm I_{421}^{\mathrm{sat}}$, which corresponds to the maximum available laser power of \SI{70}{\milli\watt} per TC beam in our experiment. To achieve optimal laser cooling, we use a TC beam detuning of $\mbox{-1}$~$\mathrm\Gamma_{421}$, which has been determined independently with the same method. With the TC setup, we are able to increase the atomic flux in the main chamber by a factor of up to 6.5, which seems to be limited only by the available laser power, since we do not observe any significant saturation of atom numbers towards high TC beam intensities.

\subsection{Zeeman Slower \label{sec:zs}}

\begin{figure}
  \includegraphics{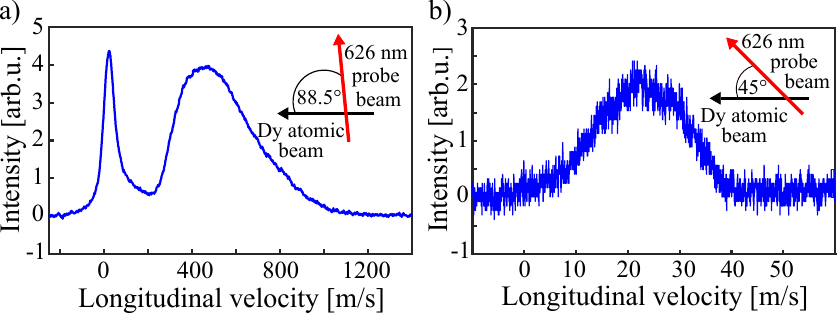}
\caption{Doppler spectroscopy of the atomic beam for two different scenarios. (a) For an angle of $\mathrm\alpha = \SI{88.5}{\degree}$ between the atomic beam and the probe beam, we measure the full velocity distribution of the atomic beam coming out of the ZS, which shows a clear signature of the slowed atoms at $\sim$\SI[per-mode=symbol]{24}{\meter\per\second}. (b) We decrease the angle to $\mathrm\alpha = \SI{45}{\degree}$ in order to measure the slow velocity part of the distribution with higher resolution.}
\label{fig5:zs_spec} 

  \includegraphics{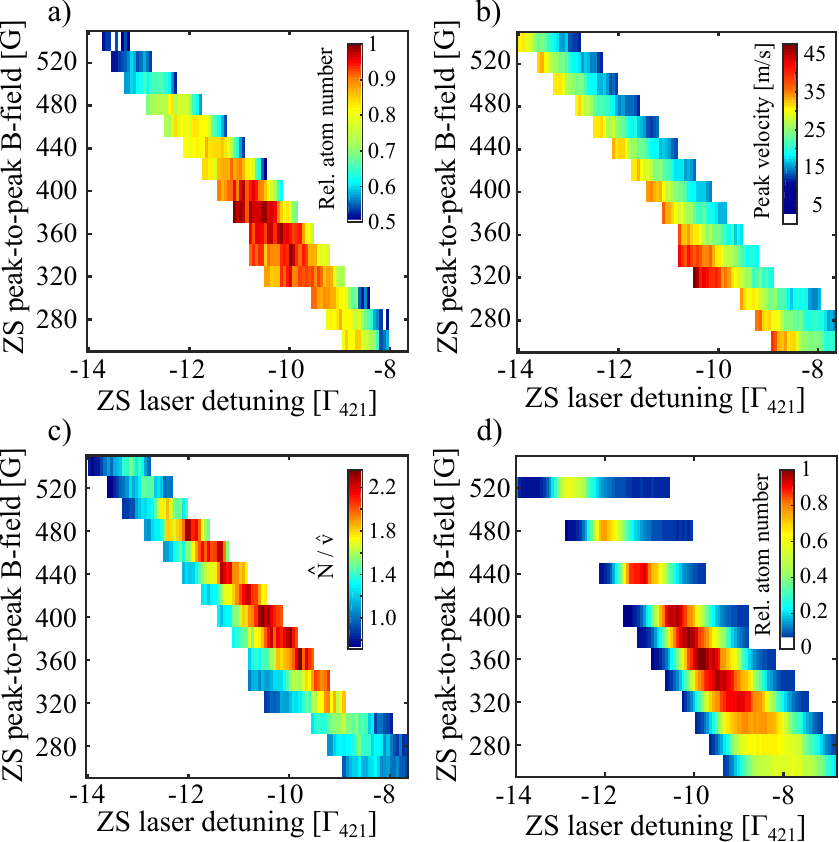}
\caption{ZS optimization measurements. (a) Map of slow atom numbers, normalized to the maximum value measured in the investigated parameter space. (b) Corresponding peak velocity map. For the determination of optimal ZS parameters with regard to MOT loading efficiency both the slow atom number and the peak velocity have to be taken into account. (c) Combination of atom number and velocity data to weight the amount of atoms with their velocity $\hat{N}/\hat{v}$. (d) Relative atom number map as a function of the ZS peak-to-peak magnetic field strength and ZS laser detuning determined by MOT fluorescence imaging.}
\label{fig6:zs_opt} 
\end{figure}

We characterize the ZS performance by means of \linebreak Doppler spectroscopy as described in Sec. \ref{sec:method} and shown in Fig. \ref{fig5:zs_spec}. We adjust angles of $\alpha=\SI{88.5}{\degree}$ between the atomic beam and the \SI{626}{\nano\meter} probe beam in order to measure the full velocity distribution of the atomic beam (see Fig. \ref{fig5:zs_spec}a). One can clearly distinguish between the signal of thermal atoms peaked at about \SI[per-mode=symbol]{480}{\meter\per\second} and the fraction of laser cooled atoms at $\sim$ \SI[per-mode=symbol]{24}{\meter\per\second}. We decrease $\alpha$ to \SI{45}{\degree} to be able to detect the fluorescence signal of the Zeeman-slowed atoms with better resolution (see Fig. \ref{fig5:zs_spec}b). Due to the large Doppler shift and Doppler broadening of the thermal atoms, their signal is well separated from the slow velocity part of the distribution, which is why we use the \SI{45}{\degree} spectroscopy setup to optimize the ZS parameters. To yield slow atom flux, the magnetic field and the laser detuning have to match the ZS resonance condition \cite{phillips1982}. In order to determine the best operating range for our ZS, we measure slow atom spectra (as shown in Fig. \ref{fig5:zs_spec}b) for different combinations of ZS currents, i.e. different ZS peak-to-peak magnetic field strengths (as depicted in Fig. \ref{fig3:mag_fields}a), and ZS laser detunings. From these spectra we determine the peak velocity and the area below the peak, which we take as a measure for the atom number. In this way, we investigate a parameter space between \SI{260}{\gauss} and \SI{540}{\gauss} of ZS peak-to-peak magnetic field strength and $\mbox{-8}$~$\mathrm\Gamma_{421}$ to $\mbox{-14}$~$\mathrm\Gamma_{421}$ of ZS laser detuning. We constrain our measurements to regions where the peak velocity of the distribution stays below \SI[per-mode=symbol]{50}{\meter\per\second}, which is well beyond the capture range of the MOT. Figure \ref{fig6:zs_opt}a shows a map of relative atom numbers, which is normalized to the maximum atom number measured within the investigated parameter space. It shows a broad maximum for ZS peak-to-peak magnetic field strengths around \SI{360}{\gauss}. However, for the determination of optimal ZS parameters with regard to the MOT loading efficiency, both the amount of atoms and their velocity distribution have to be taken into account. Figure \ref{fig6:zs_opt}b shows a peak velocity map of the corresponding slow atom distributions. As can be seen, for given ZS peak-to-peak magnetic field strengths (horizontal trends), the peak velocities vary over a considerable range, which allows to further confine the ZS parameters to regions, where the peak velocity is low. To account for both criteria, we choose a representation (see Fig. \ref{fig6:zs_opt}c), in which we plot $\hat{N}/\hat{v}$, where $\hat{N}$ denotes the relative atom number (as shown in Fig. \ref{fig6:zs_opt}a) and $\hat{v}$ represents the corresponding peak velocities, normalized to the highest velocity measured. In order to obtain a preferably high number of atoms at low peak velocities we chose to operate the ZS at a peak-to-peak magnetic field strength of \SI{380}{\gauss}, which corresponds to a current of \SI{9.5}{\ampere}, and a laser detuning of $\mbox{-10}$~$\mathrm\Gamma_{421}$.

To confirm whether the ZS parameter optimization, as described above, also accounts for optimal MOT loading, we have repeated the optimization procedure by MOT fluorescence imaging. Therefore, we load the MOT for different ZS parameters and detect the \SI{626}{\nano\meter} fluorescence light with a CCD camera after \SI{3}{\second} of loading. Figure \ref{fig6:zs_opt}d shows a map of the MOT fluorescence signal, normalized to the maximum value measured. The trend agrees with the previous atomic beam spectroscopy measurements and also shows a broad range of ZS settings that lead to almost equally high atom numbers in the MOT. Our previously determined ZS parameters of a peak-to-peak magnetic field strength of \SI{380}{\gauss} and $\mbox{-10}$~$\mathrm\Gamma_{421}$ fit very well with the MOT fluorescence optimization measurements, where we measure maximum atom numbers for ZS peak-to-peak magnetic field strengths in the range between \SI{340}{\gauss} and \SI{400}{\gauss} and ZS laser detunings between $\mbox{-9}$ and $\mbox{-11}$~$\mathrm\Gamma_{421}$. We thus find that in our setup we are able to optimize the ZS parameters individually either by fluorescence spectroscopy of the atomic beam or by fluorescence imaging and atom number optimization of the MOT.

\subsection{Magneto-optical trap \label{sec:mot}}

\begin{figure}
  \includegraphics{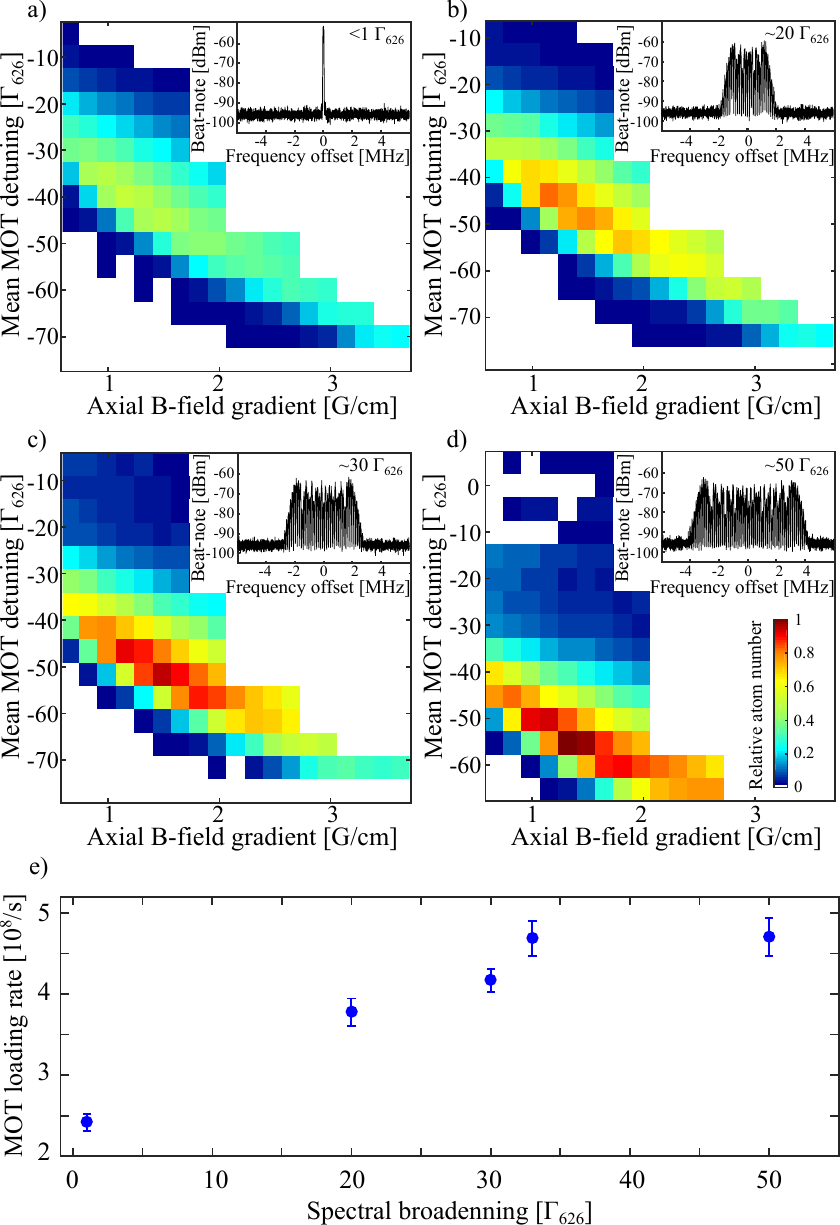}
\caption{MOT loading phase optimization. The plots show relative atom number maps for four different settings of spectral broadening as a function of the mean MOT laser detuning and the axial magnetic field gradient, measured after \SI{3}{\second} of loading. The insets show beat-note measurements of the MOT laser for the respective setting of spectral broadening. All four plots share the same color code for the atom numbers, which have been normalized to the overall maximum atom number. (a) No broadening. (b) 20~$\mathrm\Gamma_{626}$. (c) 30~$\mathrm\Gamma_{626}$. (d) 50~$\mathrm\Gamma_{626}$. (e) Comparison of MOT loading rates for different effective laser linewidths, measured with optimized parameters.}
\label{fig7:mot_load_opt} 
\end{figure}

\subsubsection{MOT loading:} During the loading phase of the MOT, we artificially broaden the \SI{626}{\nano\meter} laser linewidth via sideband modulation inside an AOM \cite{kuwamoto1999,dorscher2013}. To this end, we modulate the AOM drive signal sinusoidally with a frequency of \SI{136}{\kilo\hertz}. Depending upon the modulation amplitude, this generates a comb-like frequency spectrum around the central laser frequency of variable width, which we characterize in a beat-note detection scheme with an unmodulated laser beam (see Fig. \ref{fig7:mot_load_opt} insets). In this way, we are able to increase the effective laser linewidth from a few tens of Kilohertz up to \SI{6.8}{\mega\hertz}. We investigate its influence on the loading behavior of our MOT by determining the trapped atom number after a constant loading time of 3 seconds for different effective laser linewidths. For each setting we investigate a parameter space between axial magnetic field gradients of \SI[per-mode=symbol]{0.7}{\gauss\per\centi\meter} to \SI[per-mode=symbol]{3.5}{\gauss\per\centi\meter} and mean laser detunings of $\mbox{-10}$~$\mathrm\Gamma_{626}$ to $\mbox{-75}$~$\mathrm\Gamma_{626}$. Here, the mean laser detuning denotes the frequency difference between the center frequency of the spectrally broadened laser light and the atomic resonance, which we can adjust via the fiber EOM while keeping the spectral broadening at a constant level. We use \SI{120}{\milli\watt} per MOT beam, which corresponds to a mean intensity of 735~$\mathrm I_{626}^{\mathrm{sat}}$, in order to saturate all additional frequency components. Figure \ref{fig7:mot_load_opt} shows atom number plots, normalized to the overall maximum atom number, for four different effective laser linewidths (no broadening, 20~$\mathrm\Gamma_{626}$, 30~$\mathrm\Gamma_{626}$ and 50~$\mathrm\Gamma_{626}$, see Fig. \ref{fig7:mot_load_opt}a-d). As the linewidth is increased, more atoms can be captured, when the mean laser detuning is increased as well to keep all additional frequency components red detuned from the atomic resonance. By measuring loading curves for MOT parameters that lead to maximum atom numbers for different effective laser linewidths, we observe that the loading rates begin to saturate at an effective laser linewidth of about 33~$\mathrm\Gamma_{626}$ (see Fig. \ref{fig7:mot_load_opt}e). We use this setting for the loading phase of the MOT with a mean detuning of $\mbox{-52}$~$\mathrm\Gamma_{626}$ and a magnetic field gradient of \SI[per-mode=symbol]{1.5}{\gauss\per\centi\meter}.

\begin{figure}
  \includegraphics{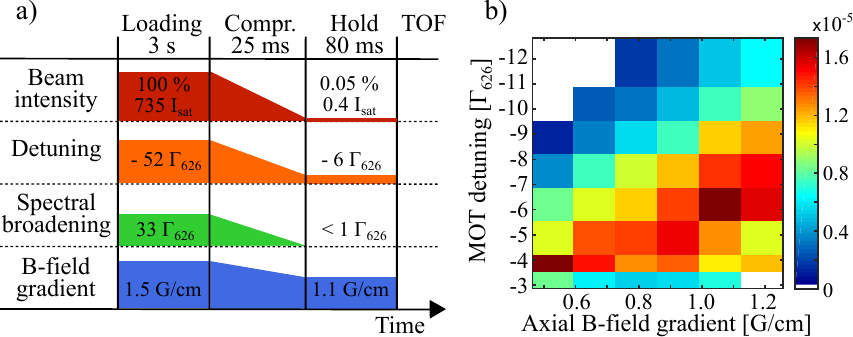}
\caption{MOT compression phase optimization. (a) Schematic drawing of the experimental sequence. After \SI{3}{\second} of loading, we compress the MOT within \SI{25}{\milli\second} to reduce the temperature and to increase the density. The atoms are held for \SI{80}{\milli\second} before we start the TOF sequence. (b) Phase space density map as function of the final axial magnetic field gradient and the final MOT detuning at the end of the compression phase.}
\label{fig8:mot_compr_opt} 
\end{figure}

\subsubsection{MOT compression:} To increase the density and to reduce the temperature of the atomic cloud, we subsequently compress the MOT by simultaneously ramping down the laser beam intensity, the mean laser detuning, the magnetic field gradient and the MOT light broadening in a linear way. Afterwards, we hold the atoms in the trap for \SI{80}{\milli\second} to let them equilibrate before we take time-of-flight absorption images. By varying the compression time between \SI{25}{\milli\second} and \SI{400}{\milli\second}, we observe that twice as many atoms are lost for \SI{400}{\milli\second} compared to \SI{25}{\milli\second}, while the final temperature and the density do not change significantly. For this reason, we use \SI{25}{\milli\second} in the compression phase. We also investigate the influence of the laser detuning and the magnetic field gradient after the compression ramp on the atom number, the temperature and the phase space density. As a consequence, we load the MOT for \SI{3}{\second} and vary the final laser detuning and the final magnetic field gradient at the end of the compression ramp in a parameter space between $\mbox{-3}$~$\mathrm\Gamma_{626}$ and $\mbox{-12}$~$\mathrm\Gamma_{626}$ and \SI[per-mode=symbol]{0.5}{\gauss\per\centi\meter} to \SI[per-mode=symbol]{1.2}{\gauss\per\centi\meter}. We observe that within the investigated parameter space the atom number predominantly exceeds $9\cdot 10^8$ and the temperatures are lower than \SI{12}{\micro\kelvin}, without major differences. However, when considering the phase space density \cite{townsend1995}, one can clearly see a distinct maximum at $\mbox{-6}$~$\mathrm\Gamma_{626}$ and \SI[per-mode=symbol]{1.1}{\gauss\per\centi\meter} (see Fig. \ref{fig8:mot_compr_opt}b). \\
Figure \ref{fig8:mot_compr_opt}a shows a schematic of our optimized experimental sequence for MOT loading and compression. After a MOT loading time of 3 seconds we simultaneously reduce the mean laser beam intensity from 735~$\mathrm I_{626}^{\mathrm{sat}}$ to 0.4~$\mathrm I_{626}^{\mathrm{sat}}$, the mean laser detuning from $\mbox{-52}$~$\mathrm\Gamma_{626}$ to $\mbox{-6}$~$\mathrm\Gamma_{626}$, the spectral broadening width from 33~$\mathrm\Gamma_{626}$ to the laser linewidth and the magnetic field gradient from \SI[per-mode=symbol]{1.5}{\gauss\per\centi\meter} to \SI[per-mode=symbol]{1.1}{\gauss\per\centi\meter}. With this optimized sequence, we are able to trap $10^9$ dysprosium atoms at a temperature of \SI{9}{\micro\kelvin} and a phase space density of $1.7\cdot 10^{-5}$, which constitutes a good starting point for transferring the atoms into an optical dipole trap and for subsequent forced evaporative cooling.

\section{Conclusion \label{sec:concl}}

We have presented our apparatus for cooling and trapping of neutral dysprosium and we have shown extensive parameter optimization measurements for the Zeeman slower and the magneto-optical trap. We have discussed, how we can optimize the ZS performance by two independent methods, which we find in good agreement with one another. For our $^{162}$Dy MOT, we have investigated the influence of artificial laser linewidth broadening on the atom number and have determined optimal parameters for the MOT loading phase. Finally, we have studied the achievable atom numbers, temperatures and phase space densities of the compressed MOT and with optimized parameters we are able to trap $10^9$ dysprosium atoms at a temperature of \SI{9}{\micro\kelvin} and a phase space density of $1.7\cdot 10^{-5}$. We believe that our results can be helpful for other experiments with cold lanthanide atoms, especially during the building phase.

\begin{acknowledgments}
We gratefully acknowledge financial support by the JGU-Startup funding, DFG-Grossger\"at INST 247/818-1 FUGG and the Graduate School of Excellence MAINZ.
\end{acknowledgments}


\end{document}